\documentstyle[twocolumn,epsf,aps,tabularx]{revtex}
\draft
\begin{document}
\title{A three-state cyclic voter model extended with Potts energy }

\author{Gy\"orgy Szab\'o and Attila Szolnoki}

\address{
Research Institute for Technical Physics and Materials Science \\
P. O. Box 49, H-1525 Budapest, Hungary}

\address{\em \today}

\address {
\centering {
\medskip \em
\begin{minipage}{15.4cm}
{}\qquad The cyclically dominated voter model on a square is extended by
taking into consideration the variation of Potts energy during the nearest
neighbor invasions. We have investigated the effect of surface tension
on the self-organizing patterns maintained by the cyclic invasions.
A geometrical analysis is also developed to study the
three-color patterns. These investigations indicate clearly that in
the ``voter model'' limit the pattern evolution is governed by the
loop creation due to the overhanging during the interfacial roughening.
Conversely, in the presence of surface tension the evolution is governed
by spiral formation whose geometrical parameters depend on the strength
of cyclic dominance.
\end{minipage}
}}

\maketitle

\narrowtext

\section{INTRODUCTION}
\label{sec:intro}

In systems with several species (particles, opinions, etc.) 
the cyclic invasion processes can maintain a self-organizing domain
structure. Although this phenomenon is investigated extensively in
different areas, such as the chemical reactions on crystal
surfaces \cite{FB,CH}, biological (Lotka-Volterra)
systems \cite{JMS82,LV,FKB,BFK}, Rock-Scissors-Paper (RSP) games in
evolutionary game theories \cite{HS}, cyclically dominated voter
models \cite{TI,T94,SSM}, the mechanism sustaining the polydomain
patterns is not well understood. At the same time, this type of 
spatial self-organizations is believed to play crucial role in
the biological evolution \cite{JMS79} and it can provide protection
for the participiants against some external invadors \cite{BH,SC}.

One of the simplest model exhibiting variations in a self-organizing
domain structure was introduced by Tainaka and Itoh \cite{TI,T94}.
In this cyclically dominated voter model
three states ($A$, $B$, and $C$) are permitted on the sites of a square
lattice. The system evolution is governed by the iteration of invasions
between two randomly chosen nearest neighbors. The cyclically symmetric 
invasion rates are characterized by a single control parameter $P$.
For example, the pair $AB$ transforms into $AA$ with a probabilty
$1/2+P$ or to $BB$ with probability $1/2-P$ ($0 \leq P \leq 1/2$).
For $P=0$ this system is equivalent to the voter model \cite{CS,HL}
exhibiting growing domains whose correlation length is proportional
to $\sqrt{t}$ if the system is started from a random initial state. 
Contrary, for $P=1/2$ $A$ beats $B$ beats $C$ beats $A$ as it happens
in the RSP game. In this former case the above rules sustain a
self-organizing, three-color polydomain structure in which the linear
domain size can be characterized by a correlation length of $\xi \simeq
2.5$ (measured in lattice unite). Tainaka and Itoh have
shown that the typical domain size diverges when $P \to 0$ \cite{TI}.
Their numerical analysis is focused on the density of vortices which
are defined by those points of a three-color map where the three
states (domain walls) meet. In fact, one can distinguish vortices
and antivortices (rotating clockwise and anti-clockwise) and they
are created and annihilated in pairs during the evolution of
domain structure. According to the early Monte Carlo (MC) simulations
the vortex density can be approximated by a power law behavior in the
limit $P \to 0$ \cite{TI}.

In the above system the domain walls separating two homogeneous
domains are very irregular. This irregularity prevents the observation
of expected spirals to be formed by the rotating vortex arms for smooth
interfaces \cite{T94,SS}. In the absence of cyclic invasion (and related 
vortex rotation) some features of the rough interfaces are already
studied by several authors considering the two-state voter
model \cite{CS,HL,CG,OMS,DG,DCCH}. In the presence of cyclic invasion,
however, the topological and geometrical features of domain structure 
are not yet investigated rigorously even for smooth interfaces.
Very recently some geometrical features and drift of a single spiral
wave are studied by using a continuous reaction-diffusion
model \cite{Sandstede,Kheowan}. Now we report a model which exhibits
a transition between different self-organizing patterns involving
those where smooth, rotating spiral arms can be observed.

The present work is devoted to study the effect of surface tension on
the self-organizing domain structures maintained by cyclic invasions
on a square lattice.
For this purpose we have introduced a model where the nearest neighbor
invasion is affected by an interfacial (Potts) energy \cite{Potts,Wu}
whose strength is controlled by a parameter $K$. Using MC simulations
we have investigated the vortex density in the stationary state for
different values of $P$ and $K$. In order to have a deeper and more
quantitative insight into the domain structure we have developed a
method to study some geometrical features of the interfaces. 
This analysis confirms the necessity of these types of sophisticated
descriptions.

\section{THE MODEL}
\label{sec:model}

We consider a square lattice where at each site ${\bf x}=(i,j)$
($i$ and $j$ are integers) there is a state variable with three
possible states, namely $s_x=0$, 1, 2. 
The Potts energy for a configuration $s=\{ s_x \}$ is defined as
\begin{equation}
H = - \sum_{<x,y>}  [ \delta (s_x,s_y) - 1]
\label{PottsH}
\end{equation}
where the summation runs over the nearest neighbor sites and
$\delta(s,s^{\prime})$ indicates the Kronecker's delta \cite{Potts,Wu}.
Notice that the coupling constant is chosen to be energy unit.
In the present form the Potts energy measures the length of equivalent
interfaces (in lattice unit $a=1$) separating the three
types of domains and its inverse estimates the average domain
radius \cite{ole}. Evidently, in the threefold degenerated
(homogeneous) ground state $H=0$.

The configuration evolves in time according to elementary invasions
between two nearest neighbor sites (${\bf x}$ and ${\bf y}$) chosen randomly.
More precisely, a pair of neighboring state variables $(s_x,s_y)$
(assuming $s_x \ne s_y$) transforms into $(s_y,s_y))$ with a
probability
\begin{equation}
\Gamma [(s_x,s_y) \to (s_y,s_y)] = {1 \over 1 + \exp{(K \Delta H +
P D )}}
\label{eq:gamma}
\end{equation}
where
\begin{equation}
\Delta H = H_{\rm f}-H_{\rm i} \nonumber
\end{equation}
is the energy difference between the final and initial states, and
$K$, as an inverse temperature, controls the effect of Potts energy
on this single site flip.
The second term in the argument of exponential function describes
the cyclic dominance with a strength $P$, where
\begin{equation}
D = \cases{+1 & if $s_x=(s_y+1)\; \mbox{mod}\; 3$ , \cr
           -1 & if $s_x=(s_y+2)\; \mbox{mod}\; 3$ . \cr}
\nonumber
\end{equation}
In the case $K=0$ the present model is equivalent to those introduced
by Tainaka and Itoh \cite{TI}. Evidently, the three-state voter model
is reproduced if $P=0$ and $K=0$. For $K>0$, however,
the interfaces become more smooth because the present dynamics
suppresses those elementary processes which increase the interfacial
(Potts) energy.

We have to emphasize that for $P=0$ this system exhibits domain growth 
independently of the value of $K$. Apparently, for $P=0$
Eq.~(\ref{eq:gamma}) satisfies the condition of detailed balance at a
temperature $1/K$ as defined for the kinetic Potts model (on the
analogy of kinetic Ising model) 
that undergoes a symmetry breaking phase transition when increasing $K$.
There exists, however, a relevant difference due to the constraint of
invasion dynamics. Namely, in the present model the new state at a given
site should be equivalent to one of the neigboring one. This means that
the changes are localized along the boundaries separating the homogeneous
domains.

Notice that the above rules conserve the cyclic symmetry among the three
states. As a result in a sufficiently large system the three states
are present with the same probability ($1/3$). For small system, however,
one of the species can extinct due to the effect of fluctuations and
finally the system evolves toward one of the three (homogeneous) absorbing
states. Henceforth our analyses will be restricted to the large system
limit that is provided by choosing the system size to be much larger
than any length characteristic to the corresponding pattern.

\begin{figure}
\centerline{\epsfxsize=7.5cm
            \epsfbox{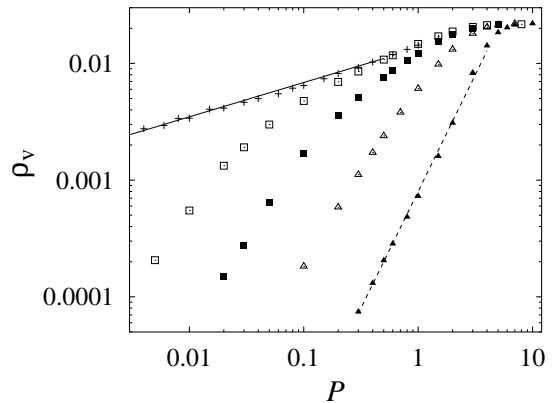}
            \vspace*{1mm} }
\caption{Vortex densities as a function of $P$ for
$K=1$ (closed triangles), 1/4 (open triangles), 1/16 (closed squares),
1/64 (open squares), and $0$ (pluses). The solid line shows the
predicted power law behavior if $K=0$. The dashed line (with a slope
of 2.05) indicates the best power law fit for $K=1$.}
\label{fig:cv_p}
\end{figure}

The above model is investigated by MC simulations under periodic boundary
conditions on a square lattice consisting of $N=L \times L$ sites. 
The system is started from a random initial state where the three states
are present with equal probabilities. During the simulations we have
recorded the number of vortices and antivortices defined above. For this
purpose we have counted those $2 \times 2$ block configurations containing
all the three possible states \cite{TI,T94,SSM}. After a suitable
transition time we have determined the average
vortex density as well as its fluctuation defined in \cite{SSM}.
The system size is varied from $L=400$ to 2000 to have sufficiently
large number of vortices in the stationary states.
The numerical results of vortex densities are summarized in a
log-log plot as demonstrated in Fig.~\ref{fig:cv_p}.

If $P>>\max (K,1)$ then the dynamics is governed by the deterministic
RSP rule that maintain a self-organizing state with small domain
sizes ($\xi \simeq 2.5$) as mentioned above. Consequently the
vortex density becomes independent of $K$ for sufficiently large
values of $P$ as demonstrated in Fig.~\ref{fig:cv_p}.

In the case $K=0$ the $P$-dependence of vortex density can be well
described by a power law, namely, $\rho_v \simeq P^{\beta}$ within
the range $0.003< P < 0.3$. The best fit is found for $\beta=0.29(1)$
confirming our previous result \cite{SSM}. At the same time 
Fig.~\ref{fig:cv_p} demonstrates clearly that
the vortex density is dramatically reduced when the interfacial
energy is switched on. For $K=1$ the MC data can be well approximated
by another power law with an exponent $\beta=2.05(9)$. It should be
emphasized that within the statistical error our data is consistent
with a quadratic behavior. Similar behavior can be conjectured 
from the trends represented by MC data for lower $K$ in 
Fig.~\ref{fig:cv_p}. Unfortunately, we couldn't confirm this
expectation by determining the leading term in the $P$-dependence
of vortex density for lower $K$ values because this analysis requires
extremely long run time and large sytems. Just to indicate the
difficulties, the determination of a data point at low vortex
densities has required more than four-week run time on a fast PC.
We think that further numerical analyses are necessary to
justify (or modify) the above conjecture.

\begin{figure}
\centerline{\epsfxsize=7.5cm
            \epsfbox{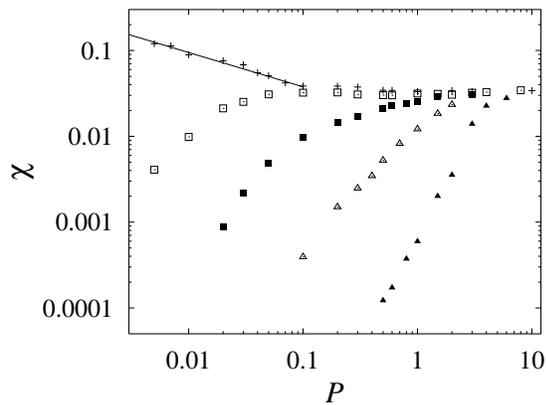}
            \vspace*{1mm}       }
\caption{Fluctuation of vortex density versus $P$ for different
$K$ values denoted by the same symbols as in Fig.~\ref{fig:cv_p}.
The solid line indicates the fitted power law divergency in the
absence of interfacial energy.}
\label{fig:cvchi_p}
\end{figure}

Due to the long run times we could derive the vortex density
fluctuations ($\chi$) with an adequate accuracy. As demonstrated
in Fig.~\ref{fig:cvchi_p} the numerical data indicate the divergency
of the vortex density fluctuation in the absence of interfacial
energy ($K=0$). Within the investigated region, this fluctuation
can be approximated as $\chi \sim P^{-\gamma}$ with $\gamma = 0.3(1)$
in good agreement with a previous result \cite{SSM}.

In the presence of interfacial energy ($K>0$) the vortex density
fluctuation vanishes with $P$ as indicated in Fig.~\ref{fig:cvchi_p}.
The vortex density fluctuation seems to be proportional to the
vortex density for sufficiently low $\rho_v$. Similar features
have been found for some systems of particles and antiparticles 
performing branching annihilating random walks \cite{SS,SSM}.
This is the reason why we have reinvestigated a parallel drawn between
the vortex dynamics in present model and a system of particles
and antiparticles as suggested in \cite{SSM}.
According to a simple idea the rotating vortices form spirals whose
long and narrow arms enhance the probability of the creation of a
new vortex-antivortex pair. The movement of vortices can be well
approximated by a random walk on a lattice. Furthermore, a
vortex and antivortex annihilate each other when meeting at the
same site during their random walks. The balance between the
annihilation process and pair creation yields an average concentration
in the stationary state. 
Within this framework a simply mean-field analysis (details are given
in \cite{SSM}) predicts that a quadratic behavior ($\rho_v \simeq P^2$)
can be reproduced if the pair creation is proportional to
$P \rho_v^{3/2}$ or $P^2 \rho_v$. From the view-point of vortex dynamics
both possibilities demand a better understanding about the relationship
between the creation of vortex-antivortex pairs and the geometry of
interfaces. 

\begin{figure}
\centerline{\epsfxsize=7.5cm
            \epsfbox{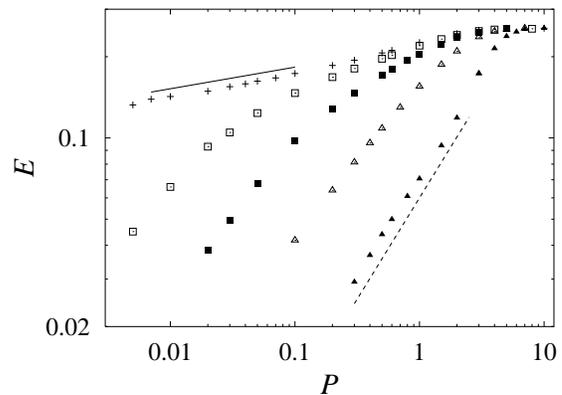}
            \vspace*{1mm} }
\caption{Average Potts energy per sites as a function of $P$ for the same
values of $K$ plotted in Fig.~\ref{fig:cv_p}. The solid and dashed lines
indicate the slopes of 0.08 ($K=0$) and 1.3 ($K=1$).}
\label{fig:PottsE}
\end{figure}

As mentioned above the whole length of interfaces is equivalent to
the Potts energy defined by Eq. (\ref{eq:PottsE}). 
During the simulations we have determined the expected value of
Potts energy per sites,
\begin{equation}
E = {1 \over N} \langle H \rangle
\label{eq:PottsE}
\end{equation}
where $\langle \cdots \rangle$ indicates the average over the sampling
time. The results of our simulations are summarized in a log-log plot
(see Fig.~\ref{fig:PottsE}).

At the first glance the $P$- and $K$-dependences of the vortex density and
Potts energy seem to be very similar. However, the detailed numerical
analysis gives different values for the exponents when fitting a power
law ($E=a P^{\alpha}$) for small $P$ values. Namely, we have obtained
$\alpha=0.08(2)$ and 1.3(1) for $K=0$ and 1 respectively. The reader can
easily check that the traditional scaling argument (predicting
$\beta=2\alpha$) is not valid in the present cases. 

\begin{figure}
\centerline{\epsfxsize=7cm
            \epsfbox{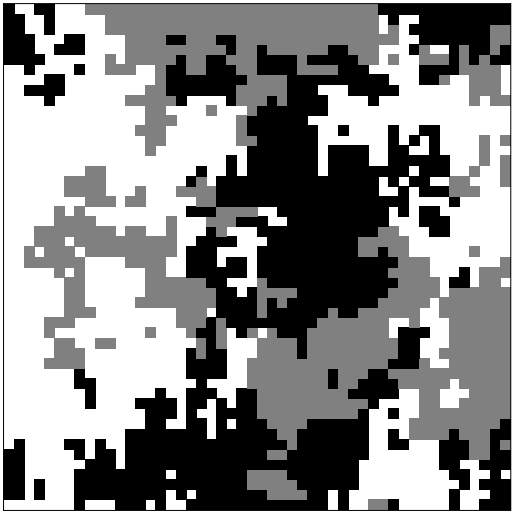}
            \vspace*{1mm} }
\caption{Typical part ($50 \times 50$) of snapshoot in a larger system
for $K=0$ and $P=0.01$.}
\label{fig:pvss1}
\end{figure}

\begin{figure}
\centerline{\epsfxsize=7cm
            \epsfbox{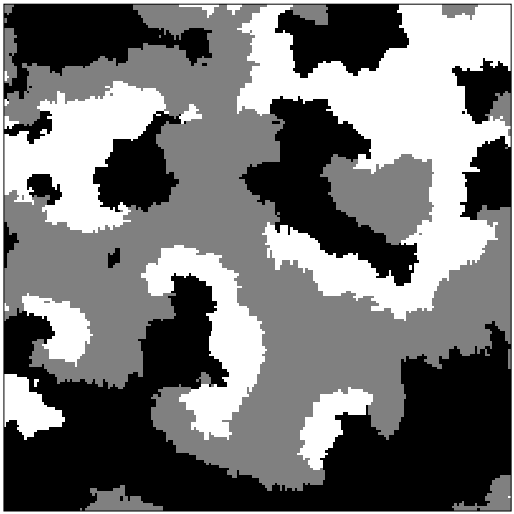}
            \vspace*{1mm} }
\caption{Spiral formation sustained by the rotating vortices and
antivortices is recognizable on a $200 \times 200$ part of a larger
system for $K=1/2$ and $P=0.3$.}
\label{fig:pvss2}
\end{figure}

The self-organizing domain structure shows striking differences depending
on wether the interfacial energy is switched on or not. For the sake of
illustration two typical patterns are shown in  Figs.~\ref{fig:pvss1}
and \ref{fig:pvss2}.
In the absence of interfacial energy ($K=0$) the nearest neighbor
invasions yield irregular boundaries whose overhanging results in small
islands (loops). Their random motion, extension, shrinking, splitting,
and fision seem to play crucial role in the pattern evolution as well as
for the three-state voter model. In this former case ($P=0$), however,
these elementary events are not able to prevent the growth of domains
whose characteristic linear size increases with time as $\sqrt{t}$
\cite{DG}.

Choosing a particular initial state it is already demonstrated that
the cyclic dominance drives the vortex rotation (for $P>0$) which
is accompaned with spiral formation \cite{T94,SSM}. In 
Figure~\ref{fig:pvss1} the rotating spirals are not recognizable
due to the irregular interfaces. However, the spiral formation 
becomes visible when the interfacial roughness is reduced by the
surface tension as demonstrated in Fig.~\ref{fig:pvss2}.

In Figure~\ref{fig:pvss2} one can easily identify the vortices and
antivortices rotating clockwise and anticlockwise respectively.  This
rotation creates spirals because the average invasion velocity is constant.
We have to emphasize that this pattern can not be characterized by a
single length unit (e.g.\ correlation length) because the main
features of spirals (armlength, average curvature, average distance,
etc.) depend on the model parameters. This is the reason why we have
developed a method to study some geometrical fetures of three-color
maps on a square lattice.

\section{GEOMETRICAL ANALYSES}
\label{sec:geometry}

On a three-color, continuous, planar map the domains are separated
by three types of smooth boundaries. Dedicated points are the 
vertices where three or more boundaries meet. If such a map evolves
smoothly then  the appearance of vertices with more than three edges
becomes negligible. Thus our analysis can be restricted to those maps
which contain only three-edge vortices and antivortices.
However, as we show later, the qualitative feature of the system 
remains uneffected if four-leg vertices are not ignored.
One can easily check that these vortices and antivortices are
positoned alternately along domain boundaries \cite{SSM}.
Our geometrical analysis will be focused on determining the average
value of arclength, rotation of tangent vector, and curvature for
those boundaries connecting a vortex and an antivortex. 
Henceforth the rotating vertex is called vortex.

On a square lattice the boundaries are polygons consisting of unit
length parts whose tangential rotation may be $\Delta \phi = \pm \pi /2$
and $0$. For a given vortex edge the tangential rotation
is determined by summarizing these quantities step by step along the edge
from a vortex to the connected antivortex. At the same time arclength
is also obtained as the number of steps.
The elementary step is based on the identification of the $2 \times 2$
block configurations. This algorithm assumes that first we have 
determined the vortex positions. To reduce the statistical error this
procedure was repeated many times during the simulations.

The above algorithm is well defined if the three-color pattern is free
of four-edge vertices. Unfortunately, the investigated self-organizing
patterns contain undesired four-edge vertices
(see Figs.~\ref{fig:pvss1} and \ref{fig:pvss2}). 
Some of them (involving all the three states) can be
considered as a vortex-antivortex pair just before their annihilation or
after they creation \cite{SSM}. The others involve only two states and 
make the paths (from vortex to antivortex) indefinite. Both types can
be removed by executing an invasion through one of the randomly choosen
four edges. Before the geometrical analysis all the investigated
distributions are slightly adjusted by repeating the random
invasions at the four-edge vertices until they vanish.
Evidently, the effect of these modifications on the energy or vortex
density is negligible if the typical domains are sufficiently large. 
The most relevant effect appears at $K=0$ when the density of
vortex-antivortex pairs is approximately $\rho_v/6$ in the whole region
of $P$ where we studied the system. Consequently, the $P$-dependence of 
vortex density remains power law after the pattern adjustment.

After having removed the four-edge vertices the pattern becomes
topologically equivalent to the continuous, three-color map mentioned
above. In this case we can distinguish two types of boundaries,
namely, loops (surrounding an isolated domain) and vortex edges
(starting at a vortex and ending at one of the connected antivortices).
Using the mentioned algorithm we have determined the average
length $l_{\rm av}$ and tangential rotation $\phi_{\rm av}$ of vortex
edges.

\begin{figure}
\centerline{\epsfxsize=7.5cm
            \epsfbox{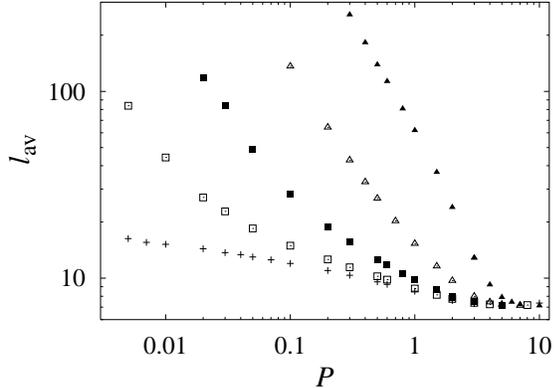}
            \vspace*{1mm}       }
\caption{Average arclength of vortex edges versus $P$ for different 
$K$ values. Symbols as in Fig.~\ref{fig:cv_p}.}
\label{fig:l_p}
\end{figure}

In Figure~\ref{fig:l_p} the log-log plot of the average length of vortex
edges shows that $l_{\rm av}$ increases slowly when $P$ is decreased
for $K=0$. Significantly faster increase can be observed in the presence
of surface tension. The arrangement of MC data for $K=1$ has inspired us
to fit a power law, $l_{\rm av}=a P^{-\lambda}$ as we had done for the
vortex density and the Potts energy per sites. Within the same region
of $P$ the best fit is found for $\lambda = 1.75(5)$.

\begin{figure}
\centerline{\epsfxsize=8cm
            \epsfbox{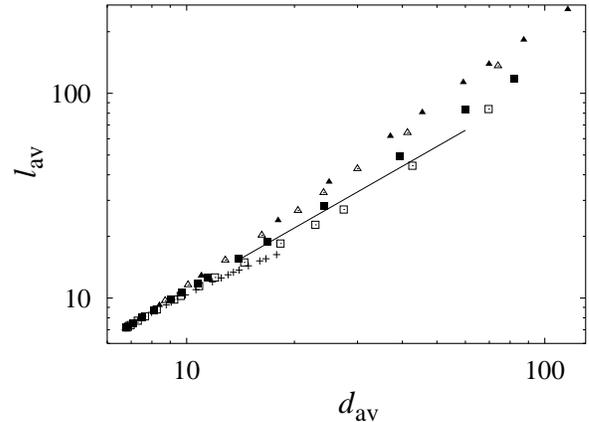}
            \vspace*{1mm}       }
\caption{Relation between the average vortex distance ($d_{\rm av}$)
and arclength of vortex edges ($l_{\rm av}$) for those parameters 
(and symbols) plotted in Fig.~\ref{fig:cv_p}. Both lengths
are measured in unit of lattice constant.}
\label{fig:l_d}
\end{figure}

The above behavior is not surprising because the average 
vortex distance exhibits qualitatively similar $P$-dependence. For
the quantitative analysis an average vortex distance can be deduced from
the density of vortices as $d_{\rm av}=1/\sqrt{\rho_v}$. The striking
difference caused by the introduction surface tension and cyclic
invasion becomes visible when the pairs of data $d_{\rm av}$ and 
$l_{\rm av}$ are plotted on a log-log plot (see Fig.~\ref{fig:l_d}).
In Figure~\ref{fig:l_d} the straight line ($l_{\rm av}=1.05 d_{\rm av}$)
demonstrates those set of domain structures that can be well characterized
by a single length scale. For example, such a situation can be observed
when considering the domain growth in the three-state Potts model below
the critical temperature. In the absence of interfacial energy the MC data
indicates significantly different relation which may be approximated
as $l_{\rm av} \simeq d_{\rm av}^{0.8}$ within the given region. The
slower increase of the average length of vortex edges can be explained
by the increasing number of those vortex-antivortex pairs which have two
common (short) edges. Shuch a pair is frequently created when the moving
islands meet the third type of domain during their random movements.

\begin{figure}
\centerline{\epsfxsize=7.5cm
            \epsfbox{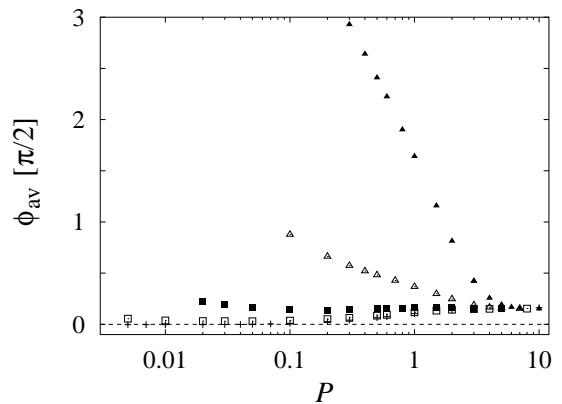}
            \vspace*{1mm}       }
\caption{Average tangential rotation of vortex edges as a function of
cyclic dominance ($P$) for different $K$ values (symbols as in 
Fig.~\ref{fig:cv_p}).}
\label{fig:phi_p}
\end{figure}

An opposite tendency can be recognized for those cases where the spiral
formation becomes relevant because the arclength of spiral
arms always exceed the distance between the corresponding vortex
and antivortex. At the same time our data reflects that the average
tangential rotation of a vortex edge increases with $d_{\rm av}$.
This statement is supported by those data in Fig.~\ref{fig:phi_p}
we obtained for $K>0$. Particularly, for $K=1$ one can observe 
that both $\phi_{\rm av}$ (see Fig.~\ref{fig:phi_p}) and $l_{\rm av}$
(Fig.~\ref{fig:l_p}) increases monotonously when $P$ goes to zero. 
Unfortunately, we were not able to study what happens when the average
tangential rotation becomes larger than $2\pi$.

In Figure~\ref{fig:phi_p} the angle of tangential rotation is measured
in unit $\pi/2$ which is a natural choice on a square lattice. 
One can observe that $\phi_{\rm av}$ becomes
practically zero in the $P \to 0$ limit in the absence of interfacial
energy.
In the light of this result one can think that the spiral formation
does not play a dominant role in the pattern formation for $K=0$. At
the same time, we should keep in mind that the cyclic invasion ($P>0$)
is required to sustain the self-organizing domain structures, otherwise
the domains would grow unlimited. Unfortunately, we cannot explain
quantitatively the microscopic mechanism yielding this behavior. Now
we can only give some additional arguments supporting the crucial
role of islands as mentioned above.

\begin{figure}
\centerline{\epsfxsize=7.5cm
            \epsfbox{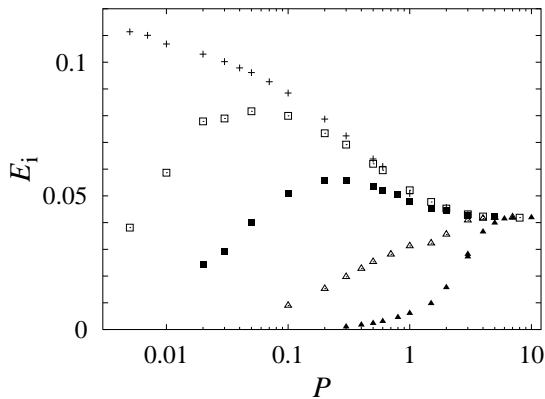}
            \vspace*{1mm}       }
\caption{The interfacial energies of islands vs. $P$  for different
$K$ values. Symbols as in Fig.~\ref{fig:cv_p}}
\label{fig:island}
\end{figure}

The total interfacial (Potts) energy can be separated into two parts. 
The first contribution comes from the island boundaries and the second
part from the vortex edges. Thus the energy per sites can be written
in the following form:
\begin{equation}
E = E_i + 3 \rho_v l_{\rm av}
\label{eq:Ei}
\end{equation}
where $E_i$ denotes the contributions of islands to the total Potts
energy $E$ defined by Eq.~(\ref{eq:PottsE}). The second term indicates
that the contribution of vortex edges can be expressed as a product
of the density of (three-edge) vortices ($\rho_v$) and the average length
of vortex edges ($l_{\rm av}$). Using this expression we can determine
the values of $E_i$ from those data plotted in Figs. \ref{fig:cv_p},
\ref{fig:PottsE}, and \ref{fig:l_p}. The results of this calculation
are illustrated in Fig.~\ref{fig:island}.

For $K=1$ the interfacial energy contribution of islands vanishes at
sufficiently low values of $P$. This
tendency can be observed in Fig.~\ref{fig:island} for other $K>0$
values. The lower the value of $K$, the lower the value of $P$ is where
$E_i$ becomes negligible. This means that in the stationary state 
the number of islands are reduced by the interfacial energy. This
tendency  can be visually checked if the reader compares the two
snapshots shown in Figs. \ref{fig:pvss1} and \ref{fig:pvss2}.

Notice, furthermore, that for $K=1$ the energy contribution of islands
is negligible for weak cyclic dominance where the $P$-dependence of $E$,
$\rho_{\rm v}$, and $l_{\rm av}$ can be well approximated by power
laws as mentioned above. The substitution of the corresponding
expressions into (\ref{eq:Ei}) yields a relation between the exponents,
namely $\alpha = \beta - \lambda$. Our numerical data support this result.   

The investigation of $E_i$ for $K=0$ shows surprising results. In the
case of strong cyclic dominance the dominant part of interfacial
energy comes from the vortex edges. According to our simulations 
the contribution of $E_i$ to the total interfacial energy
increases meanwhile $E$ decrease (see Fig.~\ref{fig:PottsE} when
decreasing $P$ in the investigated region. Since $E>E_i$ therefore
these opposite tendencies imply the possibility of a break point 
for $P<0.002$.

\section{CONCLUSIONS}
\label{sec:conc}

We have studied numericaly the effect of surface tension on the
self-organizing patterns maintained by cyclic invasions among three
species on a square latice. For this purpose the cyclic voter model
introduced by Tainaka and Itoh \cite{TI} is extended in a way conserving
the cyclic symmetries. In the original model the invasion between two
(randomly chosen) nearest naighbors is not affected by the neighborhood.
In the extended model
the nearest neighbor invasion rate is influenced by the neighborhood
via taking the variation of Potts energy into account. Our
analyses are restricted to those situations ($K \ge 0$) where this
modification favorizes those invasions which reduce the length of
interfaces separating the domains.

Our simulations have justified that the introduction of interfacial
energy causes relevant changes in the observed patterns. To have a
more quantitative and sophisticated picture we have determined the
average value of some geometrical features of the interfaces
(e.g. arclength and tangential rotation of vortex edges). This 
method is based on the analogy to the
continuous limit of a three-color map. By this way 
we could study the contributions of vortex edges and islands
separately. In the light of this analysis we can distinguish three
types of typical domain structures. 

In the deterministic limit ($P>>\max (K,1)$) the pattern consists of
small compact domains and contains many vortices and antivortices.
In the absence of interfacial energy ($K=0$) the typical domain size
as well as the contribution of island interfacial energy increases 
when the the cyclic dominance ($P$) is decreased. Here the island
creation via the interfacial roughening seems to be a relevant
phenomenon. Conversely, in the presence of interfacial energy the
islands vanish when $P$ is decreased and vortex (spiral) rotations
dominate the pattern evolution. The transitions among these
typical behaviors are smooth.

Our numerical results are obtained in a limited region of the
parameter $P$ due to the technical difficulties appearing for large
typical domain sizes. For some cases ($K=0$ and 1) our data can be
approximated by power laws in a region of $P$. We are, however, not
convinced that these (expected universal) behaviors remain valid for
lower $P$ values. For example, we don't know what happens when the
average tangential rotation of vortex edges becomes significantly
larger than $2 \pi$. Deviations can also appear for $K=0$ at lower
$P$ values where $E_i$ is expected to decrease monotonously with $P$. 

The suggested geometrical analyses confirm that the self-organizing
patterns cannot be characterized by a single length unit as it
happens for many other systems. In these cases two patterns
can not be transformed into each other by choosing a suitable length
scale. In the presence of interfacial energy this feature is strongly
related to the appearance of spiral vortex edges whose average
tangential rotation remains unchanged during such a geometrical
magnification. We think that this type of geometrical analysis joints
different approaches and models, furthermore, it motivates a
theoretical effort to find general relations among these
quantities.

\acknowledgements

We thank Tibor Antal for helpful comments. Support from the Hungarian
National Research Fund (Grant Nos. T-33098 and T-30499) is acknowledged.
A. S. thanks the HAS for financial support by a Bolyai scholarship.

\end{document}